\documentclass[12pt]{article}

\def\B{B}

\def\X{X}

\def\pa{\partial}

\def\nn{\nonu}

\def\bea{\begin{eqnarray}}
\def\eea{\end{eqnarray}}

\def\nn{\nonu}

\def\be{\begin{equation}}
\def\ee{\end{equation}}

\let\nonu=\nonumber

\def\V{{\cal V}}

\begin{document}
\bibliographystyle{perso}

\begin{titlepage}
\null \vskip -0.6cm
\hfill PAR--LPTHE P03--13

 \hfill RUNHETC-2003-06

\hfill hep-th/0304221

\vskip 1.4truecm
\begin{center}
\obeylines

        {\Large
Gravitational Topological Quantum Field Theory 
Versus $ N=2 \ D=8 $ Supergravity and its lift to $N=1\ D=11$ Supergravity
}
\vskip 6mm
{ Laurent Baulieu 
  Laboratoire de Physique Th\'eorique et Hautes Energies,
 Universit\'e  de  Paris~VI~and ~Paris~VII, France$^{ \dag}$\\
{  and}\\
 Department  of Physics, Rutgers University, USA$^{\dag\dag}$ 
  }

\end{center}

\vskip 13mm

\noindent{\bf Abstract}: In a previous work, it was shown that the
8-dimensional topological quantum field theory for a metric and a
Kalb--Ramond 2-form gauge field determines $N=1$ $D=8$ supergravity.  It is
shown here that, the combination of this TQFT with that of a 3-form
determines $N=2 $ $D=8$ supergravity, that is, an untruncated dimensional
reduction of $N=1$ $D=11$ supergravity. Our construction holds for
8-dimensional manifolds with $ Spin(7) \subset SO(8) $ holonomy.  We
suggest that the origin of local Poincar\'e supersymmetry is the
gravitational topological symmetry.  We indicate a mechanism for the lift of
the TQFT in higher dimensions, which generates Chern--Simons couplings.\vfill

 \hrule \medskip
Postal address:

$^{\dag}$Laboratoire de Physique Th\'eorique et des Hautes Energies,
 Unit\'e Mixte de Recherche CNRS 7589,
 Universit\'e Pierre et Marie Curie, bo\^\i te postale 126.
4, place Jussieu, F--75252 PARIS Cedex 05.

 $^{\dag\dag}$ Dept. of Physics, Rutgers University,
New~Brunswick,~NJ~60637,~USA

\end{titlepage}
\def\w{\wedge}
\def\o{\omega}
\def\t{\tilde}

\def\B{B}
\def\TQFT{ Topological Field Theory}
\section{Introduction}

 Poincar\'e supersymmetry and  topological  
supersymmetry are deeply related on manifolds with special holonomies.
This  was observed      in  \cite{bakasi} and \cite{acharya}, where 
 the  8-dimensional Yang--Mills  topological quantum field theory   (TQFT) was
constructed.     \cite{bakasi} also suggested    
that   
 the   introduction of a TQFT  for a  
3-form  in eight dimensions is  of relevance, for a  better understanding
 of  supergravities  and  for   determining effective
actions for the $M$   theory. 

 \cite{green} indicated the existence of a   
formal link between   topological sigma-models  and all  anomaly-free   
superstrings. For us, it was a signal   that  the origin of 
local  Poincar\'e  supersymmetry might be  the gravitational   
topological symmetry.  This raises the question whether      
$D=11$ supergravity, which   determines all known
supergravities in lower dimensions as   limits of superstrings, 
can   be  constructed in  the context of a TQFT. Eventually, there is the
possibility  that  there is a unique topological  symmetry that can be represented
in 
  two possible phases, one which is purely topological and the other one
 which describes  particles.

To concretely understand how supergravity can be deduced  from topological
gravity is not a trivial task. We  first  demonstrated  that    
topological gravities in four  and eight dimensions can be  respectively
untwisted into  $N=2
$  $D=4$   supergravity \cite{BT} and   a truncated version of 
 $N=1 $  $D=8$   supergravity \cite{BT1}. Then, we proved  that the TQFT of a
general tensor of rank two in  eight dimensions, which is made of the
symmetric metric and the Kalb--Ramond ~2-form, determines in a twisted way
the complete  $N=1 $  $D=8$   supergravity \cite{BT3}. Introducing a  TQFT for the 2-form in
\cite{BT3} has  greatly clarified    questions opened in  our earlier work that 
  were related to the interpretation of   graviphotons and  of
  Lorentz invariance.  In all these cases, the
result is that the TQFT   can be identified with an Euclidean  
supergravity,  around the solution of a
gravitational instanton. The
oversimplified  case of $N=1$ $D=2 $ supergravity was analyzed in
\cite{ba2gr}.   Moreover, a   geometrical insight on  
topological 2-dimensional  gravitational  invariance was given in \cite{bsg2}.

  We basically found in all these works that the gravitino  is a 
    topological ghost for the  reparametrization
symmetry,   up to  
twist. This is  an appealing feature, since the gravitino   can then   be  
identified as     a curvature (in an
enlarged space) and not   as  a connection. Moreover, we find  that local
supersymmetry is      a ghost of   ghost symmetry rather than    a
gauge symmetry. This changing   point of view  eliminates many of the  puzzling
questions that are related to the elusive  construction  of  a gauge group for
the Poincar\'e local supersymmetry.     
As a matter of fact, all elements of supergravity multiplets  acquire  
robust   geometrical  definitions   within  the context    of   topological
gravity   
  coupled to TQFTs of forms.

In this paper, we  reach our earlier goal,    which was
to show that the field spectrum of $N=1$ $ D=11$  supergravity can be determined  in the
context of a 8-dimensional gravitational TQFT.

 Using a freedom  in the ghost sector of the
Kalb--Ramond 2-form,   which we had not exploited in \cite{BT3}, we   
enlarge     the ghost   of ghost symmetry of the twisted $N=1$ $ D=8$
supergravity.  This  suggests adding new degrees of freedom.   The very
natural idea is to introduce   a TQFT for a 3-form and to couple it to the TQFT of 
\cite{BT3}. A second gravitino emerges,  which  is made of  some of the  topological
ghosts of the 3-form. Remarkably, the  resulting theory   turns out to
be         
 $N=2 $  $D=8$   supergravity,  whose  classical content is
made of the metric, a 3-form, three 2-forms, seven scalars, two gravitinos and six
Majorana spinors.  This is nothing but    an  untruncated  dimensional
reduction of the spectrum 
 $N=1$ $ D=11$   supergravity, in a twisted form. 

 Our construction holds for  
8-dimensional  manifolds  with
$   Spin(7) \subset SO(8) $ holonomy.    (We could as well choose a smaller
holonomy group,  for instance
$G_2$.)   In fact,    one 
covariantly constant spinor
$\epsilon$ is needed in order to relate   forms
and      spinors  by ``twist". It also  allows one  to construct  self-duality equations, which are
generally invariant only  under the action of the holonomy group. Strictly
speaking, we must thus think of a link between the gravitational  TQFT and the
supergravity theory that we  expand  around a gravitational instanton.    The
``twist" operation is what changes the    fermions of the TQFT, which exactly
balance all contributions of bosons, modulo zero modes, into fermions, which
satisfy the physical spin--statistic relation  and can be interpreted as
particles.

 After untwisting, we recover the full
Lorentz invariance. Indeed, when we perform the untwisting  from
the     TQFT toward supergravity, the explicit dependence of the TQFT
on $\epsilon$ is absorbed in the change of variables that maps forms on
spinors. Then, we get the supergravity action in its $SO(8)$ invariant form.

An intriguing question is whether   the covariantly constant spinor,
which makes the twist possible, has a physical origin, for instance,  
as   an expectation value of some field. In the presence of a brane,   
a form may exist  in higher dimensions  with a constant flux
$\varphi$ through a hypersurface, which    can  
generate  the constant spinor $\epsilon$, by relations like 
$^t\epsilon  \gamma^a\ldots\gamma^c\   \epsilon   = \varphi^ {a\ldots
c}$. Other possibilities exist.

The paper is organized as follows. The next section is devoted to  string theory
arguments,  which   further indicate    that the determination of the complete
spectrum of the maximal supergravity in the context of gravitational gravity is
probably not coincidental.    Section 3     details our precise arguments, 
which show   the  relationship  between    the 8-dimensional gravitational  TQFT for a
metric, a 2-form and a 3-form  and  $N=2
$  $D=8$    supergravity. Section 4   sketches an argument that
   new
TQFTs seem  to exist in eight  and seven dimensions, which involve as dynamical
fields   a twisted gravitino    and forms, but not a metric. Section 5 shows a
mechanism for lifting the theory in higher dimensions and generating Chern--Simons
terms.

\section{ String theory argument}

Superstring theory   induces gravity, and is  explicitly 
 reparametrization invariant in target-space. The way target-space  supersymmetry,
and thus supergravity, emerges   is   less direct. Superconformal
invariance on the world-sheet   selects vertex operators
that describe the gravitino in the quantum field theory limit. Eventually, one
discovers   target-space supersymmetry as a symmetry transformation
between the graviton and the gravitino.  The local supersymmetry
transformations are given in an infinitesimal form. They build an open
algebra   that closes
only modulo equations of motions, which makes  their  
geometrical interpretation quite obscure.

Our suggestion  is that the algebra of supergravity transformations 
does'nt   correspond to  a gauge symmetry group. It is
basically supported by the  observation   that  the 
gravitino is a curvature rather than a gauge field, since  it can be  defined   as a
combination of    untwisted  gravitational  topological ghosts. The 
latter are  actually defined  from  the action of exterior operators upon fields with  
status of connections.  The whole gravitational TQFT  construction   
suppresses   well known difficulties  that   occur  in all attempts at building a group
out of infinitesimal  supersymmetry transformations of supergravity

 The idea that supergravity can be deduced from topological gravity  is
heuristically supported by the following    world-sheet argument, in the NSR formalism.   The string
coordinate 
$X^\mu$ has  world-sheet supersymmetric partners $\psi ^\mu$ and
$\bar\psi^\mu$, where 
$\mu$ labels the world index of target-space.  Once a conformal
structure has been chosen for the world-sheet  of  the string,  
  both   generators $Q$ and $\bar Q$ of world-sheet supersymmetry
    determine a world-sheet spinor, with:
\bea\label{fond}
Q X ^\mu= \psi  ^\mu +\ldots\quad,\quad 
\bar Q X ^\mu=\bar \psi ^\mu +\ldots
\eea
The world-sheet
spinor  $(\psi ^\mu,\bar\psi^\mu)$ has the physical spin--statistic relation on
the world-sheet, but the unphysical one in target-space.  Formally,    
Eq.~(\ref{fond}) suggests   that the  induced metric $g_{\mu\nu}$ of the 
target-space quantum field theory limit  has  a symmetry of the type: 
\bea \label{fondg}
Q g_{\mu\nu}= \psi_{\mu\nu}+\ldots\eea 
The field
 $\psi_{\mu\nu}$ clearly  looks as   a target-space
topological ghost.
Its  spin--statistic relation is   unphysical, so that   it 
   contributes negatively to the energy, with  contributions that   are opposite 
to those of the metric.  
Our present understanding of TQFTs suggests to us that the symmetry in Eq.~(\ref{fondg})  must lead us to   a   quantum field theory limit that is 
 topological gravity in target-space. Only indirectly can it lead us
to supergravity, provided     forms can be mapped on spinors in the manifold, 
 and    fermions can be extracted with the  physical spin--statistic.

As emphasized above, $\psi_{\mu\nu}$ has indeed a   precise
interpretation in target-space.    As  a topological ghost for the
topological symmetry  that is  associated to   the   reparametrization
group,    $\psi_{\mu\nu}$
is   the component of a curvature in the enlarged space that unifies the
 form-grading  of fields and their ghost number. By no mean,
  can $\psi_{\mu\nu}$  be interpreted as a connection. In the topological 
BRST framework,   
$\psi_{\mu\nu}$     transforms under   a ghost of
ghost symmetry.   The Faddeev--Popov
spinor ghost for local supersymmetry  will
be obtained by untwisting the ghosts of ghosts. However, it is
not expected that ghosts of ghosts correspond to ``infinitesimal" anticommuting
spinors, 
  which would eventually be usable 
  for giving a group structure by integration.   

The proposed idea is so general, that it should apply to all known
supergravities, and thus,   to their essential parent,  which is $N=1$
$ D=11$  supergravity. And, indeed, after having achieved the details of the
construction of the TQFT for
  a metric, a 2-form and a 3-form in 8 dimensions, we
will have the desired identification between a    topological BRST algebra
and   the infinitesimal   symmetries of $N=2 $  $D=8$   supergravity, which  is an
expression of the $N=1$ $   D=11$ supergravity. It is striking  that, in this way,
all ingredients of supergravity will be    described from   geometrical
considerations in the space of      field configurations of a metric, a
2-form and a 3-form in 8 dimensions.

Another   motivation  for the  description of
supergravity as a  gravitational TQFT is    that  Eq.~(\ref{fond})
 also  indicates that the superstring theory can be twisted in a
topological sigma-model. The operators $Q$ and $\bar Q$   can be
identified as BRST and anti-BRST operators  for  the topological sigma-model  
by shifting to zero value  the conformal weight of the superstring fields  
$\psi ^\mu$ and
$\bar
\psi ^\mu$, and by doing  a  compensating twist  that transforms 
 the world-sheet $N=1$ supergravity into 2-dimensional topological
gravity,  which  ensures the conformal anomaly-free condition.   
Formally, the twist is   a mere change of variables as   for instance
indicated in
\cite{ green}\cite{ba2gr}.   For a
topological sigma-model, the dependence  on  the  target-space
local properties becomes   very loose. This suggests to us that its 
quantum field theory limit is topological gravity instead of supergravity.
For consistency,  however, 
we must be able to  recover 
supergravity from topological gravity directly,   which is
what we will achieve  in this paper. 

The proposal  that the local details of 
target-space become a secondary notion is reinforced by     observations presented
in 
\cite{green}. There, we have shown that, for a genuine world-sheet theory,
i.e, a   pure  2d-surface theory   without   matter, but with     a rich enough
2-dimensional gravitational structure  for  enforcing the absence of a conformal
anomaly, one can  extract    target-space
coordinates from   the   world-sheet structure.
Indeed, if we take  2d-supergravity with
supersymmetry of rank larger than four,   the sum of the contributions of  all
its ghosts to the  conformal anomaly 
  vanishes and  it is  thus   inconsistent  to   introduce matter
under the form of external   string coordinates. The latter   would generate an
anomaly. On the other hand,  we have shown    that we can perform various twists
of the 2d-ghosts, while keeping 
 the conformal anomaly equal to zero  and recover all possible superstring
theories, with $N<4$ worldsheeet supersymmetry, combined with decoupled TQFTs
\cite{green}.
 In this presentation, the
physical string coordinates, i.e, the coordinates of the effective target-space, 
appear as bound states of the additional ghosts that are  initially   introduced
for  describing the extended supersymmetry on the world-sheet. The superstring
coordinates are defined in an  analogous  way.  An alternative scheme is to
replace  the    2d-supergravity of rank $N$    by the topological  
$W_N$ gravity, with the interesting limiting case of $W_\infty$ gravity, which has  
  a Lie algebra structure. Eventually, the vertex operators
that describe the fields of the limiting supergravity theory are   composites
of 2-dimensional ghost 
  fields that arise  from a pure two-dimensional geometrical structure, with no
early reference to a target-space.

 These intriguing observations  have been the
support of our idea that supersymmetry in target-space is more of a topological
origin than is usually expected.   
 The next section   is to show   in detail how  the maximal supergravity  is
actually related to a topological model of the
Donaldson--Witten~type~\cite{twist}. 


 \section{Determination of the spectrum of   N=2 D=8    supergravity}

\subsection{ The TQFT for  N=1 D=8    supergravity}
In \cite{BT3}, we have constructed the TQFT  for a  tensor of
rank 2  for   manifolds with $Spin(7)$ holonomy. We obtained that   the BRST
topological multiplet is essentially: 
\begin{eqnarray}\label{mult1}
e^a_{\mu }, B_{\mu\nu}, \sigma,  A^{(2)}_\mu,  A^{(-2)}_\mu  ,
(\Psi^{(1) a}_\mu, \bar \Psi^{(-1)ab^-}_\mu, \bar
\Psi^{(-1)}_\mu), 
 (\bar
\chi^{(-1) }_{\mu },
 \Psi^{( 1) }_{\mu\nu^-}, \chi^{( 1)} )
\end{eqnarray}

This is nothing but  the spectrum of $N=1 $  $D=8$    supergravity, up
to a twist,   which  is enabled by the existence of a 
covariantly constant spinor in the manifold. 

In more detail, we   constructed in \cite{BT3} a TQFT  for the vielbein
$e^a_\mu$,  the spin connection   $\o^{ab}_\mu$ and the Kalb--Ramond 2-form
   $\B_{\mu\nu}$. We found that the topological BRST multiplets for these fields
can be expressed  as follows:
\bea\label{twoform}
\matrix
{       &    &  &       e^a_{\mu  }         
\cr
         &     &   & \swarrow   \ \  \ \ \ \ \ \ \ \ \ \ \ \ \ \  
\cr
     &     &
  \  \ \ \ \ \  \Psi^{(1)a}_{\mu }    &  &     
   (\bar \Psi^{(-1)ab^- }_{\mu}, \bar \chi^{(-1)  }_{\mu}  )     & &   
\cr
  &     &\swarrow    \ \ \  \ \ \ \ \ \  \ \ &
&  \swarrow    \ \ \ \  \
\ \ \  \ \ \ \ 
\
\
\
\     
\cr
     &       \Phi^{(2)a}  &       & 
 \sigma,\Phi^{(0)ab^-} ,  (b^{(0) ab^-}_{\mu},  b^{(0)  }_{\mu}) & &   
 \bar\Phi^{(-2)a}   
\cr
 &     &   &    \swarrow \  \ \ \ \ \ \ \  \ \ \ \ \ \ \ \ \ \ \ \ \ \ \ \ \ \
   &    \ \ \ \ \ \     \ \ \ \ \ \     \ \ \ \ \ \   \swarrow  &
    \ \ \ \ \ \         
\cr 
   &          & 
 \chi^{(1)}, \eta ^{(1)ab^-} 
 &   &      \bar \eta ^{(-1)a }       
\cr
 } 
\eea
   
\bea        
\matrix
{       &    &  &     \o^{ab}_{\mu  } \ \ \ \ \ \ \ \ \         
\cr
         &     &   &
 \swarrow \ \   \ \ \ \  \ \ \ \ \ \ \ \   \ \ \ \ \ \ \ 
\  \ \ \
\     
\cr
     &     &  
\  \ \ \ \ \ \t \Psi^{(1)ab}_{\mu }    &  &     
    \bar{\t  \Psi}^{(-1)ab  }_{\mu}     
\cr
  &     &\swarrow    \ \ \  \ \ \ \ \ \  \ \ & \ \ \  \ \ \ \ \ \  \ \ \ \ \ \ 
 \
\
\ 
\
\ \ \ \ \  \ \  \swarrow   
\cr
     &       \t \Phi^{(2)ab} 
 &       & 
 \t\Phi^{(0)ab^-} ,  \t b^{(0) ab^-
}_{\mu}  
& &   
 \bar{\t \Phi}^{(-2)ab^+}   
\cr
 &     &   &     
   &    \ \ \ \ \ \ \ \ \ \ \ \  \swarrow 
 &  &       
\cr
   &          &   
 &   &      
\bar{\t \eta} ^{(-1)ab^+ }   
\cr
  }
\eea

\begin{eqnarray}
\let\hw=\hidewidth
\matrix
{ && &&   &&    &&  B_{\mu \nu  _{}}       \cr
   &&    &&    &&         &   \hw\swarrow\hw  \cr
   && &&   &&     \Psi^{(1)}_{\mu \nu}  &&    &&
        \bar \Psi^{(-1) }_{\mu
\nu^- }  \cr
  && &&   &  \hw\swarrow\hw &&    &&   \hw    \ \ \ \ \ \ \    \swarrow\hw  \cr
  && &&   A^{(2)}_{\mu } &&    &&      \hw(A^{(0)},  A^{(0)}_{\mu
\nu^-})\ , b^{(0)  }_{\mu \nu^-}\hw &&   &&  A^{(-2)}_{\mu } \cr
  && & \hw\swarrow\hw  &&   &&  \hw\swarrow   \ \ \hw   &&&& \hw\swarrow\hw \cr
  &&R^{(3)} &&  &&     \hw  S^{(1)},\ (\Psi^{(1)}_{\mu\nu^-}, \Psi^{(1)})
  \hw &&  &&   \hw \bar S^{(-1)},  \bar \Psi^{(-1)}_{\mu}
\hw &&      &&     \bar R^{(-3)} \cr
  &\hw\ \hw   &&  &&\hw\swarrow\ \ \ \ \ \ \hw    &&&& \hw\swarrow\ \ \ \hw 
&& &&\hw\swarrow\ \ \hw \cr
   &&  && \hw b_ {S^{(1)}}^{(2)} \hw &&  &&   b_ {\bar S^{(-1)}} ^{(0)} 
&&        &&   \hw b_{\bar R^{(-3)}}^{(-2)} \hw &&   &&    }
\label{B2}                                                                     
\end{eqnarray}
 \vskip 0.5 cm
To  gain control on the gauge invariance of the graviphoton
$A^{(-2)}_{\mu }$, we need a   BRST
quartet that includes  an Abelian  ghost
$
 d^{(-1)}$ . We thus     extend  the pair   $(A^{(-2)}_{\mu},
\Psi^{(-1)}_{\mu})$ in Eq.(\ref{B2}) as follows \footnote{The BRST
symmetry for this  U(1) invariance is expressed by: 

  \noindent $s  A^{(-2)}_{\mu } =  \bar \Psi^{(-1)}_{\mu} +\partial_\mu
d^{(-1)}$,  
 $s\bar \Psi^{(-1)}_{\mu} =- \partial_\mu   \Phi   ^{(0)}$, 
$s  d^{(-1)}=  \Phi   ^{(0)}$, $s\Phi   ^{(0)}=0$, $  
s\Phi   ^{(0)}=\bar \eta ^{(1)}$,  $ s \bar \eta ^{(1)} =0 
$.
}:
     \bea
\matrix
{          &  &  A^{(-2)}_{\mu }  &&\hookrightarrow      
\cr
            & \swarrow       
\cr
    \bar \Psi^{(-1)}_{\mu}\cr\cr \cr \cr\cr
\cr
  } 
\matrix
{          &  &    A^{(-2)}_{\mu }          
\cr
               & \ \  \ \ \ \ \ \  \  \  \ \swarrow     
\cr
           &          \bar \Psi^{(-1)}_{\mu},   d^{(-1)} 
\cr
       &\swarrow \ \ \ \ \ \ \ \ \ \  
\cr
     \ \ \ \     \Phi^{(0)}  &       &   &   \ \ \ \ \ \ \ 
\bar  \Phi^{(0)}
\cr
     &   &      & 
 \swarrow    
&       
\cr
         & &        &          
         \bar \eta^{(1)}  \ \ \ \ \ \ \ \ \ \ \ \  &
  &    
\cr
  }   
\eea

 We also have the usual ghost systems for the reparametrization and  Lorentz
invariances:
 \bea
\matrix
{
           \xi^ \mu    &     &    \bar \xi^ \mu \ \ \ \ \ 
\cr
     &   \ \ \ \ \ \  \swarrow
\cr
     &    b^\mu \ \     
  }
\quad\quad\quad\quad\quad
\matrix
{
      \Omega^ {ab  }    &   &        \bar \Omega^ {ab  } \ \ \ \ \
\cr
     &  \ \ \ \ \ \   \swarrow
\cr
     &  b^{ab  } \ \  \ \  \ \   
  } 
\eea

 In these equations, some  2-forms have been decomposed in a
Spin(7) invariant way  as 
$X_{\mu\nu }=X_{\mu\nu^-}+X_{\mu\nu^+}$, where  $X_{\mu\nu^-}$ and
 $X_{\mu\nu^+}$  are 
self-dual and antiself-dual projections of   $X_{\mu\nu}$,   with  dimensions 7 and 21
respectively, according to
$28=7\oplus 21$.   We have
slightly improved  our  general notation  of~\cite{BT3}  by
adding   
 southwest arrows,  which  indicate which fields are related by
topological BRST transformations.

 For each field, the upper index  in
parenthesis indicates the ghost number.
  The fields that  are not 
on the left edge of each pyramid come into  topological pairs that are made of a
commuting or an anticommuting   antighost  $\bar g$ with   its  Lagrange
multiplier $\lambda$  of the opposite statistics. They   satisfy  BRST equations that  can
be  basically  written as 
$
    s \bar g^{(g)} = \lambda ^{(g+1)},\   s \lambda ^{(g+1)} =0 
$, 
 i.e, that are of a trivial type. These BRST equations  can be improved  in  a way
that  expresses   the reparametrization symmetry and the
symmetries of forms  in an equivariant way. 

   The fields that  carry  the essential geometrical
information are on the left edge of each pyramid.  We refer to  ~\cite{BT3}, where
  the  explicit     BRST transformations      of  all above fields have  been
displayed in great details.

All  topological BRST
equations  actually derive from  geometrical equations on   extended curvatures,
which are of the following type:  
\bea
\label{hg}
(s+d)(B^{ }_{\mu\nu }dx^\mu \wedge dx^\nu
+ 
V^{ (1)}_{\mu  }dx^\mu  
+
m^{ (2)} )\cr
      =\exp {i_\xi} \Big(dB_2+   \Psi ^{(1)}_{\mu\nu } dx^\mu  \wedge  dx^\nu
+   A ^{(2)}_{\mu  } dx^\mu  
+R ^{(3)}\Big)  
\eea  
The operator $\exp {i_\xi}$ generically    takes  into account the
reparametrization symmetry. 

   The expansion in ghost number of Eq.(\ref{hg}) and of its   Bianchi
identity 
    defines the BRST symmetry.  Eq.(\ref{hg}) also     shows
that  the topological ghosts and ghosts of ghosts, which appear in its right
hand-side, are  components of a curvature. This turns out to be the important
observation  in view of a geometrical   interpretation of the gravitino as
curvature and not as a gauge field for a gauge symmetry.  This interpretation 
generalizes to the antighost sector, and is allowed by the existence of a
tri-grading, made of the ordinary form degree, the ghost number and the   
antighost number

%

It was    found in~\cite{BT3} that the  expression of the  topological
gravity  action is:
\begin{eqnarray}\label{top-act}
 \int _{{M_8}} \cal L &=&\int _{{M_8}} s 
\Big[ 
\bar \Psi^{(-1)ac^-} (b ^{(0)cb^-} + \o^{cb^-}(e) - G_d^{cb^-} e^d)\V_{ab}  \cr
& &\ \ +\bar\chi^{(-1)a} \Big (b^{(0)}_a +\partial_a \sigma
+\Omega_{abcd}G_{bcd}  \Big  ) \Big]
\cr 
& & + s \Big[  \partial_{[\mu}A^{(-2)}_{\nu]^-} (\Psi^{(1)a}_{[\mu}
e^a_{\nu]}+\Psi^{(1)}_{\mu\nu} ) \ \Big]\cr &&
s\Big[ 
e^\mu_c e^\nu_d \  \bar{\tilde\Phi}^{(-2)cd^+}( {
\Psi}^{(1)}_{\mu\nu} -\Psi^{(1)a}_{[\mu}e^a_{\nu]})\Big],
\end{eqnarray}
where
${\cal V}_{a_1\ldots a_i}\equiv {1\over (8-i)!} \epsilon_{a_1\ldots
a_8}e^{a_{(i+1)}}\ldots e^{a_8}$.  This topological action  gives kinetic terms
for the graviton,   the
gravitino, the 2-form $B_{\mu\nu}$, both  two graviphotons  
$A^{\pm2}_\mu$    and the dilaton and  dilatino  of the  of $N=1$ $  D=8$
supergravity, around an octonionic gravitational instanton and a given
configuration of the dilaton.      It was emphasized in
\cite{BT3} that, by suitable deformations of the topological gauge functions, it
is possible to reproduce the exact supergravity  action.  

To  gauge-fix the local supersymmetry, which
emerges as        an   invariance of the topological ghosts in the action
(\ref{top-act}),  we added in~\cite{BT3}   the following $s$-exact action:
\begin{eqnarray}
\label{gfghost}
 \int _{{M_8}} s\Big[\sqrt{g}(\bar \Phi ^{(-2)a}   D_\mu   \Psi ^{{(1)a}  }_\mu+
{\Phi}^{(0)ab^-}  D_\mu   \bar \Psi ^{(-1)ab^-}_\mu
+\bar \Phi ^{(0) } \pa_\mu   \bar \Psi^{(-1) }_\mu)\Big]
\end{eqnarray}
 This expression is quite instructive to us. It  allows us to respectively
identify    
$(\Phi^{(2)a},  {\Phi}^{(0)ab -},\Phi^{(0)})$,
$(\bar \Phi^{(-2)a},  {\t \Phi}^{(0)ab -},\bar \Phi^{(0)})$ 
and $(\bar \eta^{(-1)a},   \eta^{(1)ab ^-} , \eta^{(1)})$ 
as   twisted versions of the  Faddeev--Popov spinor  
ghost and antighost  for local N=1  supersymmetry, and of  the  fermionic   
Lagrange multipliers for the gauge-fixing of the gravitino, with  a gauge
function
$ (\gamma^\mu  \partial_\mu)  \gamma^\nu \lambda_\nu$.   

   The gauge-fixing of both graviphotons
$A^{\pm2}_\mu$  was obtained from:
\bea
 \int _{{M_8}}
s(S^{1}\partial_\mu A^{-2}_\mu +R^{-3}\partial_\mu A^{ 2}_\mu).
\eea
Thus,   $(d^{1},S^{-1})$ and
$(R^{-3},R^{3})$ can be identified as the Faddeev--Popov ghosts and antighosts   
for the $U(1)$ invariances of both graviphotons.

The mapping between the fermionic degrees of
freedom between the fermionic sector of the TQFT and of   D=8, N=1 supergravity 
uses  the    covariantly constant spinor   $\varepsilon$  of  the  manifold
with $Spin(7)$ holonomy.  Call respectively   $ (\lambda ,
\bar\lambda)
$ and
 $ (\chi, \bar\chi) $
     the chiral and antichiral parts   of the gravitino
and   dilatino. Their  relation with    the topological ghosts 
of the TQFT  was found to be  \footnote{ The 8-dimensional gamma matrices 
$\gamma_a$ act on spinors of definite chirality. We have defined 
$\chi_{\mu\nu^-} = \Psi_{\mu\nu^-}^{(1)}-    \Psi^{(1)a}_{[\mu} e^a_{\nu]^-} $  
and $\chi_{ab^-} =e_a^\mu e_b^\nu \chi_{\mu\nu} $
\cite{BT3}.}:
 \begin{eqnarray}
\lambda &=& \Psi^a \gamma_a \varepsilon \ \ , \nonumber\\ 
\bar\lambda &=& \bar\Psi \varepsilon + 
\bar\Psi^{ab^-}\gamma_{ab}\varepsilon\ ,
\label{twist-1}  \nonumber\\ 
\chi &=& \bar\chi^a \gamma^a \varepsilon \ \ , \nn\\
\bar\chi &=& \chi\varepsilon +
\chi^{ab^-}\gamma_{ab}\varepsilon \ \ .
\label{twist-2}
\end{eqnarray}

 \subsection{Completion of the spectrum of  N=1  D=8   supergravity  by the
TQFT of a 3-form}

The previous section   section was for  recalling   the link
between  $N=1 $  $D=8$   supergravity  and  a TQFT that     involves the 
fields in Eq.~(\ref{mult1}).  We now show the novel result that   the construction
of a TQFT for a 3-form and its coupling to the TQFT that we have just written,
determines 
     $N=2 $  $D=8$   supergravity.  

 In order to reach  the $N=1 $  $D=8$  
supergravity, 
  we have   set   to zero  in a
BRST invariant way the fields
$A^{(0)}_{\mu\nu^-}$ and $A^{(0)}$, which    appear in the TQFT multiplet  in
Eq.~(\ref{B2}). We used a term
$s \Big[ 
{\bar{ \Psi}} ^{(-1 ) }_{\mu\nu ^- } { {A}} ^{(0 ) }_{\mu\nu^-}
+{\bar S}^{(-1)}A^{(0)}
\Big]$  \cite{BT3}. Here,  we will  use our freedom of   
possibly 
  defining a propagation for these fields, by a change of the topological gauge
functions, and reach eventually the $N=2 $  $D=8$   supergravity.

 We thus  go back  to
the situation where the degrees of freedom in
$A^{(0)}_{\mu\nu^-}$ and
$A^{(0)}$ are arranged  as the  elements of a vector ghost of ghost
$A^{(0)}_{\mu }$.   The latter will  be    interpreted shortly
 as the  twisted expression  of   the chiral part of a commuting
spinor, which turns out to be  the     Faddeev--Popov
antighost  for  the second generator of N=2 supersymmetry. This suggests that we need  
new fields, for representing this symmetry.
The natural idea is to introduce a 3-form gauge field
$C_{\mu\nu\rho}$, together with its topological multiplet,   and to complete the N=1 theory
into the N=2 theory, by addition of    a TQFT for the 3-form. 

We now must find the way  to write the BRST multiplet for   a  3-form.     
Following  the  notation  that is analogous to  that we have used  for the BRST
multiplet of the 2-form
$B_{\mu\nu }$, the 
  BRST multiplet of 
 the 3-form  $ C_{\mu \nu \rho _{}} $ is:
 \begin{eqnarray}
\let\hw=\hidewidth
\matrix
{ && &&   &&    &&  C_{\mu \nu \rho _{}}      \cr
   &&    &&    &&         &   \hw\swarrow\hw  \cr
   && &&   &&     \psi^{(1) }_{\mu \nu \rho } &&    &&
          \bar \psi^{(-1) }_{\mu \nu\rho\sigma^+} \cr
  && &&   &  \hw\swarrow\hw &&    &&   \hw\swarrow\hw  \cr
  && &&   B^{(2) }_{\mu \nu } &&    &&      \hw B^{(0) }_{\mu \nu ^-},
  b^{(0) }_{\mu \nu\rho\sigma^+}\hw &&   && B^{(-2) }_{\mu \nu } \cr
  && & \hw\swarrow\hw  &&   &&  \hw\swarrow\hw   &&&& \hw\swarrow\hw \cr
  &&R^{(3) }_{\mu  } &&  &&     \hw   S^{(1) }_{\mu  }, \psi^{(1)}_{\mu\nu^-}
  \hw &&  &&   \hw S^{(-1) }_{\mu  },\psi^{(-1)}_{\mu\nu}
\hw &&      &&     R^{(-3) }_{\mu  } \cr
  &\hw\swarrow\hw   &&  &&\hw\swarrow\hw    &&&& \hw\swarrow\hw 
&& &&\hw\swarrow\hw \cr
M^{(4)}  &&  && \hw N^{(2) } ,     b^{(2) }_\mu\hw &&  &&   N^{(0) } , b^{(0)}_\mu
&&        &&   \hw N^{(-2) }  , b^{(-2) }_\mu \hw &&   && M^{(-4) }   \cr
&&   & \hw\swarrow\hw & && & \hw\swarrow\hw
&&&  & \hw\swarrow\hw & && & \hw\swarrow\hw \cr
&&\eta^{(3) }  &&  &&        \eta^{(1) } && &&  \eta^{(-1) } &&  &&
  \eta^{(-3) }
}
 \label{threeform}
\end{eqnarray}
 
The  BRST equations for all these fields can be obtained in an analogous way as
  for those the
2-form, with equations as in  Eq.(\ref{hg}). 

In  Eq.(\ref{threeform})  the
topological antighost of the 3-form has been  already identified  as a  self-dual
4-form
$\bar \psi _{\mu\nu\rho\sigma^+}$. This can be accepted as an input.
Alternatively, we can start from an earlier stage and  define    the 56 components
of the topological antighost of
$C_3$ in a $Spin(7)$-invariant way as
$  35\oplus 21$.  Then, we can eliminate the 
$21$-component   against the component
    $\B^{(0)}_{\mu\nu^+}$ of the  ghost of ghost   $\B^{(0)}_{\mu\nu }$, with a
term of the form 
$s(  \bar \psi^{(-1)}_{\mu\nu^+} \B^{(0)}_{\mu\nu^+}) $. Eventually,  
     this     explains   the absence   of  the  forms $\bar
\psi^{(-1)}_{\mu\nu^+}$,
$\B^{(0)}_{\mu\nu^+}$, $b^{(0)}_{\mu\nu^+}$ and 
$\psi^{(1)}_{\mu\nu^+}$  among the fields that we have  displayed  in  the table
(\ref{threeform}).  This reduction is natural, since we basically need  a 
self-dual  Lagrange multiplier 4-form 
 $ b^{(0) }_{\mu
\nu\rho\sigma^+}$  for
determining the square of the 4-form curvature $G_4=dC_3$ of the 3-form
$C_3$,  using  a standard self-duality gauge function $G_4=^*G_4$, and
  $\int_{M^8}|G_4^+|^2= \int_{M^8} (|G_4|^2 +G_4\wedge G_4)$.

As a generalization of  what we did  for    the vector  ghost
of ghost of the Kalb--Ramond 2-form, we must  introduce the
fields that are
  relevant   to gain    control of    the gauge invariance of the  
2-form  ghost
of ghost
$\B^{(\ -2)}_{\mu \nu }$.  We thus     
replace  the field 
$ 
\bar
\Psi^{(-1)}_{\mu\nu} $ within Eq.~(\ref{threeform}) by the  following  more refined set of fields:
\bea
\matrix
{           \psi^{(-1)}_{\mu\nu}  & & \hookrightarrow       
\cr \cr \cr\cr\cr 
} \ \ \ \ \ \ 
\matrix
{           &   &     \psi^{(-1)}_{\mu\nu} , \bar
R^{(-1)}_{\mu }           
\cr
               &     \swarrow   
\cr
                   \Phi^{(0)}_{\mu} &  &
&       
      & \bar\Phi^{(0)}_{\mu}    
\cr 
             &  &  &\swarrow \
     \cr&              &  
  \eta^{(1)}_\mu   
\cr
      } 
\eea
where 
\bea\label{surprise}
\matrix
{            \Phi^{(0)}_{\mu }          
  & & \hookrightarrow       
\cr \cr \cr\cr\cr 
} \ \ \ \ \ \ 
\matrix
{           &   &  \Phi^{(0)}_{\mu } , 
m^{(0)}    &   &    &       
\cr
               &     \swarrow   
 &      &    &    &     &   
\cr
                  c^{(1)}_{ } &  &
&       
      & \bar c^{(-1)}_{ }      &
   &   
\cr
             &  &  &\swarrow \
     \cr&              &  
 \bar  \X^{(0)}    
      }
\eea
\bea
\matrix
{           \bar \Phi^{(0)}_{\mu }          
  & & \hookrightarrow       
\cr \cr \cr\cr\cr 
} \ \ \ \ \ \ 
\matrix
{           &   & \bar \Phi^{(0)}_{\mu } , 
\bar m^{(0)}    &   &    &       
\cr
               &     \swarrow   
 &      &    &    &     &   
\cr
                 \bar c^{(1)}_{ } &  &
&       
      &   c^{(-1)}_{ }      &
   &   
\cr
             &  &  &\swarrow \
     \cr&              &  
  \X^{(0)}    
\cr
      } 
\eea
\bea\label{threeform1}
\matrix
{          \eta^{( 1)}_{\mu }          
  & & \hookrightarrow       
\cr \cr \cr\cr\cr 
}
 \ \ \ \ \ \    \matrix
{           &   &    \eta^{( 1)}_{\mu } , \bar
\eta^{( 1)}_{(X) }   &   &    &       
\cr
               &     \swarrow   
 &      &    &    &     &   
\cr
                   X^{(2)}  &  &
&       
      & \bar X^{(-2)}        &
   &    
\cr
             &  &  &\swarrow  
     \cr&              &  
\eta^{( -1)}_{(X) }    
\cr
      }
\eea
 
These substitutions amount to the addition of BRST quartets  that count altogether
for zero   degrees of freedom, and can be cast in the general framework of
unification between form degrees, ghost number and antighost number. The
propagation of these fields will be  given shortly, as well as their
interpretation.

 Obtaining    the second gravitino of the theory
from the BRST multiplet of the 3-form  is  less intuitive    than      the
first gravitino. As shown in Eq.(\ref{twist-2}),
the first gravitino is a rather obvious  combination of   $Spin(7)$ covariant  
topological ghosts   of the metric and of the Kalb--Ramond 2-form. 
Our claim is that
the  chiral and antichiral components of the second gravitino are   
obtained by untwisting the following sets of forms, which come from the
topological multiplet of the 3-form:   
\bea 
&( {\bar \psi}^{(-1)} _{\mu\nu\rho\sigma^+},  \psi^{(-1)} _{\mu\nu }, 
\eta ^{(-1)}_{(X)} )
\cr
& (  \psi^{( 1)} _{\mu\nu\rho }, 
\eta ^{( 1)}_{\mu}  ) .&\label{grtwo}
\eea

Each one of these  combinations   has  64 components that are
    irreducibly decomposed   as $35\oplus 21\oplus 7\oplus 1$  under $Spin(7)$  
and as 
$56 \oplus 8  $ under $SO(8)$.
 By appropriately multiplying these tensors by gamma matrices,  and applying the
resulting $8\times 8$ matrices  on the  covariantly  constant spinor $\epsilon$ of
the
$Spin(7)$ holonomy manifold,   one gets    chiral and antichiral spinors:
 \bea
\lambda'_\mu&=& \Big(\bar \psi^{(-1)}_{\mu\nu\rho\sigma^+}\gamma^{
\nu\rho\sigma} + \psi^{(-1)} _{\mu\nu }\gamma^{ \nu } 
+\eta ^{(-1)}_{(X)}  \gamma_\mu \Big)\epsilon\cr 
\bar \lambda'_\mu&=& \Big( \psi^{( 1)} _{\mu\nu\rho }\gamma^{ \nu\rho }+
\eta ^{( 1)}_{\mu} \Big)  \epsilon
\eea
These spinors determine a  
combination of a gauge-fixed gravitino with a Higgsino.  We will be more
precise on this decomposition shortly. 

\def\B{B}

 We now start to define the Lagrangian of all these fields,  in   the tree
approximation.  Eventually, this will allow  for their physical interpretation.
We determine  the kinetic energy of the  second gravitino  of the action   by the
sum of   the following  BRST-exact terms:
\bea\label{gravitino2}
\int s\Big(\bar \psi^{(-1)}_{\mu\nu\rho\sigma^+} 
({1\over 2} b ^{(0)}_{\mu\nu\rho\sigma^+}
+
\partial _{[\mu}C_{\nu\rho\sigma]}) \Big)
&=&\int  {1\over 2} 
|\partial _{[\mu}C_{\nu\rho\sigma]}|^2 -
\bar \psi^{(-1)}_{\mu\nu\rho\sigma^+} 
\partial _{[\mu}\psi^{(1)}_{\nu\rho\sigma]}
\cr
\int s\Big( \psi^{(1)}_{\mu\nu\rho } 
\partial _{[\mu}\B^{(-2)}_{\nu\rho]} \Big)
&=&\int
\partial _{[\mu}\B^{(2)}_{\nu\rho]} 
\partial _{[\mu}\B^{(-2)}_{\nu\rho]}
-
\bar \psi^{(1)}_{\mu\nu\rho } 
\partial _{[\mu}\psi^{(-1)}_{\nu\rho]}\cr
\int s\Big(  \psi^{(-1)}_{\mu\nu  } 
\partial _{[\mu}\bar \Phi^{(0)}_{\nu ]} \Big)
&=&\int
\partial _{[\mu} \Phi^{(0)}_{\nu ]} 
\partial _{[\mu}\bar \Phi^{(0)}_{\nu ]} 
-
  \psi^{(-1)}_{\mu\nu  }  
\partial _{[\mu}  \eta^{(1)}_{\nu ]}
\cr
\int s\Big(\bar X^{(-2)}_{   } 
\partial _{\mu}  \eta^{(1)}_{\mu } \Big)
&=&\int
\bar X^{(-2)}_{   } 
\partial ^{\mu}
\partial _{\mu} X^{(2)}_{ } 
-
\eta  ^{(-1)}_{ X^2  } 
\partial _{\mu}  \eta^{(1)}_{\mu } 
\eea
The sum of the fermionic  terms is Eqs.~(\ref{gravitino2}) gives  the
Rarita--Schwinger action  for the  second gravitino, expressed in a
twisted form, and with  an algebraic gauge condition
of the type
$\gamma^\mu  \lambda_\mu=0$. The compensating   Faddeev--Popov ghost
dependent action will be determined shortly.

The bosonic terms  that are present in
the right-hand sides of Eqs.~(\ref{gravitino2}) also indicate that we are on the
right track  for determining $N=2 $  $D=8$   supergravity. They provide      
gauge invariant kinetic energy   for two 2-forms,
$\B_{\mu\nu}^{(  2)}$ and $\B_{\mu\nu}^{(- 2)}$, two graviphotons,
 $\Phi _{\mu }^{(0)}$ and  $\bar \Phi _{\mu }^{(0)}$,
and two scalars,  $\bar X _{ }^{(-2)}$ and  $X _{  }^{(2)}$.

The dependence on the 3-form $C_3$ is 
equivariant with respect to the tensor gauge invariance  
$C_3\sim C_2+d\Lambda_2$, so it is only through the
curvature $dC_3$. However,  the topological field theory BRST construction   
provides   all necessary  fields    for gauge-fixing the local
symmetries of the   topological ghosts   ghosts of ghosts of $C_3$.
%

We must   gauge-fix $\B_{\mu\nu}^{(  2)}$ and $\B_{\mu\nu}^{(- 2)}$, whose
classical action has already  been  obtained in the second line of 
Eqs.~(\ref{gravitino2}). This is done from the following term:
\bea \label{gg}
&\int  s \Big(S^{(1)}_{\mu  } 
(\partial _{ \nu}\B^{(-2)}_{\mu\nu}  
+
\partial _{ \mu}N^{(-2)}  )
+
R^{(-3)}_{\mu  } 
(\partial _{ \nu}\B^{(2)}_{\mu\nu}
+
\partial _{ \mu}N^{(2)}  )+S^{(1)}_{\mu  } b^{(-2)}_{\mu  } \Big)
\cr
& 
=\int
 \Big ( \partial _{ \nu}\B^{(-2)}_{\mu\nu}
 \partial _{ \rho}\B^{(2)}_{\mu\rho}
+
\bar N^{(-2)}_{   } 
\partial ^{\mu}
\partial _{\mu} N^{(2)}_{ } 
 \cr & \ \ \ \ \ \ \ \ \ \ \ \ \ \ \ \ \ \ \ \ 
+
 \partial _{ [\mu} S^{(1)}_{\nu  ]} 
 \partial _{ [\mu}R^{(-1)}_{ \nu]}
+
S^{(1)}_{\mu   }( \partial _{ \nu}\psi^{(-1)}_{\mu\nu} +
\partial _{ \mu}\eta^{(-1)}_{ })
\cr
&  \ \ \ \ \ \ \  
+
\partial _{ [\mu} R^{(3)}_{\nu  ]} 
 \partial _{ [\mu}R^{(-3)}_{ \nu]}
+
R^{(-3)}_{\mu   }   
\partial _{ \mu}\eta^{(3)}_{ }
\Big)
\eea
 This expression shows us
  that  the propagators of     vectors ghosts must  be gauge-fixed, so we
add the further term:
\bea\label{ggg}
&\int   s \Big(  \bar M^{(-4)}  
 \partial _{ \mu}R^{(3)}_{\mu }  
+
\bar m^{(0)}
 \partial _{ \mu}R^{(-1)}_{\mu  } \Big)
\cr
& 
=\int
 \Big (  
 \bar  M^{(-4)}_{   } 
\partial ^{\mu}
\partial _{\mu} M^{(4)}_{ } 
+
\bar  m^{(0)}_{   }  
\partial ^{\mu}
\partial _{\mu} m^{(0)}_{ }
+\bar  m^{(0)}_{   }  \partial _{ \mu}\Phi ^{(0)}_{\mu  }
 \cr &   
+
  \bar \eta^{(-3)}  
 \partial _{ \mu}R^{(3)}_{\mu }  
+
\bar c^{(1)} 
 \partial _{ \mu}R^{(-1)}_{\mu  } \Big)
\eea
We now understand   that 
$( R^{(3)}_{\mu }, R^{(-3)}_{\mu })$ and $ ( R^{(-1)}_{\mu }, S^{(1)}_{\mu })$ are 
  respectively   standard   Faddeev--Popov vector ghosts and antighosts for the
covariant gauge-fixing of   both  2-form gauge fields $
\B^{(\pm 2)}_{\mu\nu}$ \footnote{We must redefine
$\eta^{(1)}_{\mu }\to \eta^{(1)}_{\mu }+S^{(1)}_{\mu }$    to absorb the
term 
$   S^{(1)}_{\mu   }  \partial _{ \nu}\psi^{(-1)}_{\mu\nu}$ in   Eq.~(\ref{gg}).
Analogously, we  redefine    
$\bar  X^{(0)} \to \bar  X^{(0)}+ \bar  m^{(0)}$ to  absorb the  
term~$\bar  m^{(0)}_{   } 
\partial _{
\mu}\Phi ^{(0)}_{\mu  }$.}.
The   fields    
$ M^{(\pm 4)}_{ }$, $ N^{(\pm 2)}$   and $(m^{(0)}, m^{(0)}) $ are   the  standard  
   ghosts of ghosts of these  vector ghosts.  Thus, the  sum of
both actions  in Eqs.~(\ref{gg}) and  (\ref{ggg}) is nothing but the  
covariant gauge-fixing Lagrangian  for both 2-forms  $
\B^{(\pm 2)}_{\mu\nu}$ in a generalized   Feynmann gauge, including all
relevant   ghosts for the vector gauge symmetry of  
2-forms. This result shows   the efficiency  of  the TQFT construction
when it comes to   encoding   all  gauge symmetries of the fields.

A  remaining task is that of gauge-fixing the Abelian symmetries of $\bar 
\Phi^{(0)}_\mu $ 
and $\Phi^{(0)}_\mu$. This is done by writing:
\bea\label{aggg}
&\int   s \Big (  \bar c^{(-1)}  
 \partial _{ \mu}\Phi^{(0)}_{\mu }  
+
   c^{(-1)}  
 \partial _{ \mu}\bar \Phi^{(0)}_{\mu }+  \bar c^{(-1)} X^{(0)}   \Big)
\cr
&  
=\int
 \Big (  
 \bar  c^{(-1)}_{   } 
\partial ^{\mu}
\partial _{\mu} c^{(1)}_{ } 
+
  c^{(-1)}_{   }  
\partial ^{\mu}       
\partial _{\mu} \bar c^{(1)}_{ }
+\partial _{ \mu}\bar \Phi^{(0)}_{\mu }\partial _{ \nu}  \Phi^{(0)}_{\nu }
\Big)
\eea
In the last equation, $ X^{(0)}$ and $\bar X^{(0)}$ have been eliminated by Gaussian
integration. This  gives the gauge-fixing action for $\bar \Phi_\mu ^{(0)}$ and $ 
\Phi^{(0)}_\mu $ in the Feynman gauge.    $(\bar c^{(1)},  
c^{(-1)})
$ and
$(  c^{(1)},
\bar c^{(-1)})$  in Eq.~(\ref{surprise}) are thus  identified respectively as 
the   Fadeev--Popov ghosts and antighosts for the Abelian invariance of $\bar
\Phi_\mu ^{(0)}$ and $ 
\Phi_\mu ^{(0)}$.

 The  propagation of the fields
 $(\B^{(0)}_{\mu\nu^- }, N^{(0)}_{ }, b^{(0)}_{\mu }, S^{(-1)}_{\mu },
\psi^{(1)}_{\mu\nu^- } , \eta ^{(1)} )  $ of   the 3-form  TQFT multiplet
remains to be defined, which we now do.

 The fields  $(\B^{(0)}_{\mu\nu^- }, N^{(0)}_{ })$   determine   up to   twist 
the    chiral part of a  commuting Majorana spinor, which we will    identify
  as   the    Faddeev--Popov ghost of the second generator of
local supersymmetry in 8 dimensions. We   consider the action:
 \bea\label{secondsym}
&\int   s \Big (  S^{(-1)}  _\mu( 
 \partial _{ \nu}\B^{(0)}_{\mu\nu ^-}    
+
 \partial _{ \mu}N^{(0)}   ) \Big)
\cr
& 
=\int
  S^{(-1)}  _\mu( 
 \partial _{ \nu}\psi^{(1)}_{\mu\nu^- }   
+
 \partial _{ \mu}\eta ^{(1)} )
+   
b^{(0)}  _\mu( \partial _{ \nu}\B^{(0)}_{\mu\nu ^-}  
+
 \partial _{ \mu}N^{(0)})
\eea
The last
term is identified as the ghost part of  a Faddeev--Popov
action,    for a gauge condition that sets   eight  components of the second 
gravitino   equal to zero. These eight components build    the    chiral part of
the    spin one-half component of the second gravitino. 
  We are actually gradually  unveiling the second local supersymmetry.
The first term in Eq.~(\ref{secondsym}) is  a Dirac Lagrangian for the
anticommuting  Majorana
 spinor that one   obtains by untwisting   $(  S^{(-1)}_{\mu },
\psi^{(1)}_{\mu\nu^- } , \eta ^{(1)} ) $. This field is    
one  Higgsino of the supergravity.

The gauge-fixing term of the antichiral part of the second  gravitino involves, as
already suggested, the  fields $(A^{(0)}_{\mu\nu^- }, A^{(0)}_{ })\sim
A^{(0)}_{\mu  } $    of the spectrum of the 2-form $B_2$
in~Eq.(\ref{twoform}).  These fields were  set equal to zero  in
\cite{BT3}, so as to  obtain    $N=1$ $ D=8$  supergravity.  We now choose
another gauge-fixing for defining    the TQFT, and define a  propagation 
for
 these commuting ghosts,  by means of the action: 
 \bea\label{asecondsym}
&\int   s \Big (  A^{(0)}  _\mu( 
 \partial _{ \nu}\Psi^{(-1)}_{\mu\nu ^-}    
+
 \partial _{ \mu}\bar S^{(-1)}   ) \Big)
\cr
& 
=\int
  \Psi^{( 1)}  _\mu( 
 \partial _{ \nu}\psi^{(-1)}_{\mu\nu^- }   
+
 \partial _{ \mu}S ^{(-1)} )
+   
A^{(0)}  _\mu( \partial _{ \nu}b^{(0)}_{\mu\nu ^-}  
+
 \partial _{ \mu} b^{(0)})
\eea
The last term is indeed the desired Faddeev--Popov term for the antichiral part of
the second supersymmetry. The first term in  Eq.   (\ref{asecondsym}) is another
twisted Higgsino term.  The  sum  of both  second  terms in   the  actions  
(\ref{secondsym}) and  (\ref{asecondsym})   determine a complete   
Faddeev--Popov ghost action  for  a gauge condition 
$\Gamma^\mu  \lambda_\mu=0$, where  $\lambda$ stand for   the  chiral and 
antichiral      ungauged-fixed components  of the  second gravitino. We
conclude that, in a twisted form, the  Faddeev-Popov  ghosts
and antighosts for the second component of  local N=2 supersymmetry are:  
\bea 
 & (   b^{(0)}_\mu ,   \B^{(0)}_{\mu\nu ^-},  N^{(0)}  )& \cr 
&(   A^{(0)}_\mu  ,   b^{(0)}_{\mu\nu ^-},  b^{(0)}  )  &
\eea 

Of course,   one may feel      puzzled by the fact that the
TQFT     chooses different    gauge conditions for    the gravitinos  of  both
sectors of $N=2$ local supersymmetry.  This is actually explainable,
since   both gravitinos originate as topological
ghosts of different objects, a 2-form and a 3-form. This interesting inelegance
might disappear, if we are able to covariantly formulate our equations in higher
dimensions, perhaps in nine dimensions. In the untwisted model, it is 
however possible to readjust the gauge, in such a way that the symmetry of both
supersymmetry  sectors is recovered.

As for the number of degrees of freedom in the TQFT, the
gauge-fixing that   is revealed by   the actions  
in  Eqs.~(\ref{secondsym}) and 
(\ref{asecondsym}) 
suggests that only 56 components of each  one of the multiplets  in
Eq.~(\ref{grtwo})   are really part of the second gravitino. 

 We are now  in the 
position of   more precisely   identifying the fields in Eq.~(\ref{grtwo}). They
must be understood   as made of the twisted gauge-fixed  second gravitino and   
of one  independent  Majorana  spinor,
$(\eta ^{( 1)}_{\mu}, 
\psi^{(-1)} _{\mu\nu  ^-},  
\eta ^{(-1)}_{(X)} )    $. The latter will be  a  third Higgsino of the
supergravity, 
according to:
\bea
\label{submagics}
  ( {\bar \psi}^{(-1)} _{\mu\nu\rho\sigma^+},   \psi^{(-1)} _{\mu\nu  }, 
\eta ^{(-1)}_{(X)} )
 &\to&  ( { \bar  \psi}^{(-1)} _{\mu\nu\rho\sigma^+},  \psi^{(-1)} _{\mu\nu ^+
})\oplus (   \psi^{(-1)} _{\mu\nu  ^-}, 
\eta ^{(-1)}_{(X)} )   
\cr
 (  \psi^{( 1)} _{\mu\nu\rho }, 
\eta ^{( 1)}_{\mu}  )  &\to&    \psi^{( 1)} _{\mu\nu\rho }\oplus 
\eta ^{( 1)}_{\mu}    
\eea 

 We should now rest and contemplate  the field content   of the theory
that we have constructed step by step.

  The idea was to define a propagation for
all fields   of  the  BRST  multiplet of a 3-form, and to complete the TQFT 
 of \cite{BT3}. The TQFT sector stemming from  a 2-tensor  (the 
metric and the Kalb--Ramond 2-form) yields a propagating theory for the vielbein,
one scalar (the dilaton),  the 2-form, two vectors (two Abelian graviphotons), one
Majorana gravitino and one Majorana spinor, according to:
\bea
\label{twoform2}\matrix{
 &e^a_{\mu } , B_{\mu\nu} \to & \cr
 &e^a_{\mu }, B_{\mu\nu}, \sigma,  A^{(2)}_\mu,  A^{(-2)}_\mu  ,
(\Psi^{(1) a}_\mu, \bar \Psi^{(-1)ab^-}_\mu, \bar
\Psi^{(-1)}_\mu), 
 (\bar
\chi^{(-1) }_{\mu },
 \Psi^{( 1) }_{\mu\nu^-}, 
\chi^{( 1)} )&}
\eea

 The TQFT sector stemming from  the 3-form yields  a second Majorana
gravitino in a twisted form,  two Abelian 2-forms, two   vectors,
 two scalars and three
Majorana spinors, according to:   
\bea\label{threeform2}
\matrix{
&C_{\mu \nu \rho } \to
\cr
&
C_{\mu \nu \rho }, 
\B^{(\pm 2)}_{\mu \nu } ,
\bar\Phi^{(0)}_{\mu} , \Phi^{(0)}_{\mu}, X^{(2)} , \bar X^{(-2)}, 
( {\bar \psi}^{(-1)} _{\mu\nu\rho\sigma^+},   \psi^{(-1)} _{\mu\nu  ^+},
   \psi^{( 1)} _{\mu\nu\rho} ),&\cr& 
(\eta ^{( 1)}_{\mu},  \psi^{(-1)} _{\mu\nu  ^-}, 
\eta ^{(-1)}_{(X)} ),\ 
(  S^{(-1)}_{\mu },
\psi^{(1)}_{\mu\nu^- } , \eta ^{(1)} ) ,\ 
 ( \Psi^{( 1)}  _\mu, \psi^{(-1)}_{\mu\nu^- }   
,S ^{(-1)} )
  &
} \eea
The  second gravitino is obtained     in  a gauge of
the type
$\gamma_\mu
\Lambda_\mu=0$.

 The rest of the fields, which appear in the   topological multiplets
of  the 2-form and 3-form
multiplets,   play the
role of ordinary  Faddeev--Popov ghosts for gauge-fixing
 all gauge symmetries of these
fields that can be identified as classical fields  of 
$N=2$ $ D=8$ supergravity. We left aside the gauge-fixing of the 2-form
$B_{\mu\nu}$ and 3-form
$C_{\mu\nu\rho}$, for the symmetry $ C_3\sim C_3+d c_2$ and $ B_2\sim B_2+d c_1$,
which is standard in the equivariant construction, as it is already well understood
in the context of the topological Yang--Mills theory.

  Putting everything together,  we have therefore  constructed a
theory whose physical fields are, up to twist,  a metric, a 3-form, three
2-forms, four 1-forms, three  scalars, two gravitinos and four Majorana spinors.

This is not yet the complete $N=2 $  $D=8$   supergravity. However, we left aside 
the possibility of completing the TQFT multiplet of the 3-form by   topological
sets that involve  two 1-forms,  
$
\bar\Phi^{(-2)}_{\mu} $ and 
$ 
\Phi^{(2)}_{\mu}$, for gaining  control to the possible gauge symmetry of
$\B^{(0)}_{\mu
\nu }$.  The  fields  $
\bar\Phi^{( 2)}_{\mu} $  and $
\bar\Phi^{(-2)}_{\mu} $     play a role for
$\B^{(0)}_{\mu \nu }$ that   is analogous to that    played by  $
\bar\Phi^{(0)}_{\mu} $ and  $ 
\Phi^{(0)}_{\mu}$  for
$\B^{(-2)}_{\mu \nu }$ \footnote{A dissymmetry occurs however  in the TQFT,  because
the anti-self part $\B^{(0)}_{\mu \nu^+ }$ of  $\B^{(0)}_{\mu \nu }$ is
set equal to zero, while all components of $\B^{(-2 )}_{\mu
\nu }$ propagate.}. 

We introduce $
\bar\Phi^{(-2)}_{\mu} $ and 
$ 
\Phi^{(2)}_{\mu}$  as parts of topological multiplets, whose components   
count altogether  for zero degree  of freedom.
Such multiplets    are generically 
of the form    
$(A_\mu, \Psi_\mu, \eta, \kappa_{\mu\nu^-},\phi, \bar\phi)$.  They must
be completed with    Faddeev--Popov ghost and antighost   and a Lagrange
multiplier,
$(g,\bar g,h)$.  The TQFT action that encodes  the Abelian invariance is 
$\int s   (
\kappa_{\mu\nu^-}( H_{\mu\nu^-} +\partial_{[\mu}A_{\nu]^-})
+\eta \partial_{ \mu}\Psi _{\mu }
 +\bar g (\partial_\mu A_\mu+h) )$, which  a twisted form of the eight dimensional 
supersymmetric Yang--Mills theory~\cite{bakasi}.

We can therefore    incorporate in   the topological multiplet of the
3-form, which is given in Eqs~.(\ref{threeform}-\ref{threeform1}),
  both following topological submultiplets;
\bea
 \matrix
{          &  &    \Phi^{(-2)}_{\mu }          
\cr
               &  \ \ \ \ \ \ \ \   \swarrow   &   
\cr
           &             \psi^{(-1)}_{\mu},g^{(-1)}  
&  &  \chi^{( 1)}_{\mu\nu^- },\bar g^{( 1)} 
\cr
       &   \swarrow  \ \ \ \ \ \ \ \ \ \ \ \   
&      & \swarrow \ \ \ \ \ \ \ 
\cr
        \phi^{(0)}  &       &   b^{(2)}_{\mu\nu^-  } ,  h^{(
2)}   &  &
\bar  \phi^{(0)}\ \ \ 
\cr
     &   &          
&   \ \ \ \ \ \ \ \ \ \ \   \swarrow   
\cr
         & &                  
        & \bar \eta^{(1)}    &
  &    
\cr
  } \nonumber
\eea
\bea
 \matrix
{          &  &    \Phi^{( 2)}_{\mu }          
\cr
               &  \ \ \ \ \ \ \ \   \swarrow   &   
\cr
           &             \psi^{(3)}_{\mu}, g^{(3)} 
&  &  \chi^{( -3)}_{\mu\nu^- },\bar g^{( -3) } 
\cr
       &   \swarrow  \ \ \ \ \ \ \ \ \ \ \ \   
&    & \swarrow \ \ \ \ \ \ \ 
\cr
        \phi^{(4)}  &       &    
 b^{(-2)}_{\mu\nu^-  }, h^{(-2)} 
 &&
 {{\bar \phi}^{( -4)}}, &  
\cr
     &   &          
&   \ \ \ \ \ \ \ \ \ \ \   \swarrow   
\cr
         & &                  
        & \bar \eta^{(-3)}    &
  &    
\cr
  }  
\label{oneform2}
\eea

The TQFT action of 
these fields is made of two     twisted super Yang--Mills actions. We add each of   them   to
the  TQFT actions that  we have  already constructed.

 We now face  with a theory whose physical field content is made  of  
         the fields in Eq.~(\ref{twoform2}-\ref{threeform2}),   plus 
    two  
graviphotons, two Higgsinos and  four scalars, which stem from the fields
that  we just introduced in Eqs.(\ref{oneform2}). Thus, up to twist, our 
gravitational TQFT predicts the following  set  of  fields, which are defined
modulo ordinary gauge invariances:
  \bea\label{beauty}
g_{\mu\nu}, C_{\mu\nu\rho}, 3 B_{\mu\nu }, 6 A_\mu, 7S, 2\lambda, 6
\chi.
\eea
The notation is that  $S$,  
$\lambda$ and 
$\chi$ respectively denote scalar
fields,  Majorana  gravitinos and  Higgsinos.
This set of fields fields 
is  nothing,
but the spectrum  \cite{ sase} of $N=2$ $ D=8$ supergravity.  The Lagrangians of the TQFT and   of supergravity  coincide in the quadratic  approximation for the fermions.
  
All other fields that have occurred in our construction    can be understood as
conventional   ghosts and Lagrange multipliers for fixing the gauge symmetries of the
fields in Eq.~({\ref{beauty}). The way all fields transform under the TQFT symmetry has
been determined from the tri-complex structure that one associates to the gauge symmetries of
gravity coupled to a 2-form and a 3-form. The expression of the  topological BRST
transformations can be reinterpreted to   reproduce    the  symmetries of
$N=2$
$ D=8$   supergravity, including local supersymmetry.
 
The fields 
 that  are displayed in Eq.~(\ref{beauty})  can  also be
understood  as the spectrum of   N=1, D=11 supergravity. Indeed,      
the 
  straightforward  dimensional reduction  of  N=1, D=11 supergravity gives:
\bea\label{beauty11}
 (g_{\mu\nu}, C_{\mu\nu\rho},  \lambda )_{D=11}
\sim
(g_{\mu\nu}, C_{\mu\nu\rho}, 3 B_{\mu\nu }, 6 A_\mu, 7S, 2\lambda, 6
\kappa)_{D=8}    
\eea

To conclude this section, let us insist on the following points. As for the  identification 
 of the TQFT action and  the supergravity action,  our demonstration 
holds  at the quadratic level for the fermions. Suitable modifications of the gauge
functions by higher order terms should   reveal the exact laws of supergravity. Moreover,
the word identification means that the gravitational TQFT reproduces the supergravity around
a   $ Spin(7)\subset SO(8)$ invariant background. The
$SU(2)$ internal symmetry of the $N=2$ $ D=8$  supergravity multiplet is not explicit in our
construction. This is a direct consequence of the fact that, in the
8-dimensional TQFT, both gravitinos have different origins,  as  
topological ghosts of a 2-form  of a 3-form.

\section {A digression about the TQFT for the 3-form}

It is actually an interesting question to investigate whether a TQFT for
the 3-form alone has some interest, in particular for the study of
8-dimensional manifolds with $Spin(7) $ holonomies and of  7-dimensional
manifolds with $G_2$  holonomies. This was suggested as early as in \cite{bakasi}. 
We would like to  briefly sketch new ideas relative
to this question and indicate a possible hint  for getting an interacting theory for the 
3-form, when topological
gravity is decoupled.

We have seen in the previous section that the  TQFT multiplet for a
3-form   is basically made of the following fields: 
\bea
\label{gen}
C_3, 2B_2, 2A _1, 2S,
\lambda^\alpha_\mu,2\chi^\alpha 
\eea
These fields are those that   propagate in the TQFT, provided one does a
suitable gauge-fixing.
  Spinorial notations have already been
used for  the Fermi fields
that are topological ghosts.

 We understood in the previous section that the   gravitino
$\lambda^\alpha_\mu$ (a 1-form which a spinorial index)  is made of  
topological ghosts, including  the self-dual antighost
$\chi_{\mu\nu\rho\sigma^+}$ of $C_3$  The multiplet (\ref{gen}) does not involve a
metric. 
If we follow the conventional pattern for building a TQFT, we introduce an external
metric  $g_{\mu\nu}$, and do  a topological BRST invariant gauge-fixing. In this way, we   
can  get a free TQFT action for the fields in Eq.~(\ref{gen}). The action contains the
term:    
 \begin{eqnarray}
 \int d^8x \ s ( \chi^{\mu\nu\rho\sigma^+}(\partial _{[\sigma }
B_{\mu\nu\rho]}+  H_{\mu\nu\rho\sigma^+})) 
\cr =
 \int d^8x \  ( |dB_3+^*dB_3 |^2+ \ldots  )  
\cr=
 \int d^8x  \  \sqrt {g} (
g^{\mu\mu'}g^{\nu\nu'}g^{\rho\rho'}g^{\sigma\sigma'}
 \partial _{[\sigma} B_{\mu\nu\rho]}
 \partial _{[\sigma'} B_{\mu'\nu'\rho']}+\ldots ) 
\end{eqnarray}
The dots are easy to compute. We can dimensionally reduce the fields and get a TQFT  in
seven dimensions,  for a 3-form and a 2-form. The self-duality condition in eight
dimensions is replaced by the following one  in
seven dimensions,  
which also stands for     35 conditions:
\begin{eqnarray}
\label{bog}
dC_3+^*dB_2=0
\end{eqnarray}
   Imposing these conditions in a BRST-invariant
way   is    allowed by the fact that
$C_3$ and
$B_2$ carry 20 and 15 degrees of freedom modulo gauge transformations in
seven dimensions. Eq. (\ref{bog}) is a generalization of
the well-known Bogomolny equation.

If 
we can define a metric
$g_{ij}(B_3)$ in seven dimensions, which depends on the 3-form, and  use it  for
the 
$^*$
Hodge operation,   we    can trigger interactions between the
forms, by defining   the following action:
\begin{eqnarray}
 \int d^7x s ( \chi^{ijk  } (
(dB_2+^*dC_3)_{ijk  }+ H_{ijk  })\cr= 
 \int  d^7x   ( |dB_2+^*dC_3 |^2+ \ldots  ) 
\cr =
 \int d^7x  \sqrt {g}  (g^{ii'}g^{jj '}g^{kk'} 
 \ \partial _{[i } B_{jk ]} \ 
 \partial _{[i'} B_{j'k' ]} \cr+
g^{ii'}g^{jj '}g^{kk'}g^{ll'}
\ \partial _{[i } C_{jkl]}
 \partial _{[i'} C_{j'k'l']}+\ldots )
\end{eqnarray}
The dots   involve   the metric dependence  of the 3-form. 
This action is    analogous in spirit   to that of supersymmetric
mechanics in curved space. The subtlety is to find the relevant configuration space  of the 
3-form for defining the path integral.

 A further reduction in six dimensions determines   
theories with self-dual 2-forms. Such TQFTs have been directly studied in six dimensions,
and  interesting correspondences with supersymmetry have already been found
\cite{west}.

There is another topological field theory, which we can obtain by
directly starting from    the topological action $\int _{M_7} C_3\wedge
dC_3$.  For certain manifolds, one can  give
a special role to the direction $x^7$ and write:
\begin{eqnarray}
\int _{M_7} C_3\wedge dC_3=
 \int _{M_7} d^7x ( C_{7ab  }\epsilon^{ab cdef }\partial_{[c} C_{def]  }+
\epsilon^{ab cdef } C_{abc  } \partial_7 C_{def  }  )
\end{eqnarray}
Here, $a,b,c,d,e,f$ stand for 6-dimensional indices.

Generalizing  the analysis  of the Chern--Simons theory in three dimensions
\cite{cswitten},      formal manipulations suggest that the
7-dimensional theory with the   above action  might   also describe  a theory    
  in six dimensions, depending of a 2-form, with an
action   of the type:
\begin{eqnarray} 
 \int  _{M_6}  d^6x   \sqrt {g} 
\ g^{aa'}g^{bb '}g^{cc'} \ 
 \partial _{[a } B_{bc] }\ 
 \partial _{[a'} B_{b'c'] } +{\rm WZZ \ action}
\end{eqnarray}
A heuristic argument  can be done  by  formally integrating out $ C_{7ab  }$,  
solving the constraint
$ \partial _{[a } B_{bc ]}=0$, and picking a slice at $x^7=cte$. Eventually, the
6d-metric depends on the 2-form,
$g^{aa'}=g^{aa'}(dB_2)$, and we have a non trivial model.  The observables in the
seven dimensional theory  theory are obtained from the flux of the 3-form on a
3-cycle:
\begin{eqnarray}
 exp  \int \int \int _{\Gamma_3}C_{ijk} dx^i \wedge dx^j \wedge dx^k  
\end{eqnarray}
and
\begin{eqnarray}
 exp  \int \int _{\Gamma_2}B_{ij} dx^i \wedge dx^j  
\end{eqnarray}
  We leave the  investigation of these theories  for future work.

\section {Higher dimensions and N=1 D=11 supergravity}}

 We have   shown in  section 3 that the 8-dimensional gravitational TQFT
reproduces, up to twist,  the
$N=2\ D=8$ supergravity theory in the quadratic approximation in the
gravitino fields, around a $Spin(7)$ invariant vacuum. However, we  
left open the question of precisely determining the mechanism according to
which Chern-Simons interactions are built in the TQFT.

We wish  to look in more details how we can lift in higher dimensions the
8-dimensional TQFT that we have built 
in section 3. 
 
  Let us recall that $N=1\
D=11$ supergravity can be   reduced to type IIA 
10-dimensional supergravity, which, in turns, can be reduced to  a
9-dimensional model and, eventually, to 
$N=2\ D=8$ supergravity theory.  
  
  This  9-dimensional model can also be deduced from type IIB 
10-dimensional supergravity. The   9-dimensional
theory  is     interesting due the remarkable properties of   $Spin(9)$  
\cite{ramond}.

To relate the  8-dimensions  TQFT  with another one in 9
dimensions,  we  tentatively refer to the  mechanism explained in 
\cite{nikraba}. There, it is shown on general grounds that, if one has a
TQFT in
$d$ dimensions, with an action of the type ${\cal I}_{d}=\int_{M_d}(d(...)
+\{Q,...\})$, where $Q^2$ vanishes modulo some   gauge transformations,
then, it  exists another operator $Q_{d+1}$ in $d+1$ dimensions, such
that 
\begin{eqnarray}
Q_{d+1}^2= \partial}_ {  x^{d+1},
\end{eqnarray}
 modulo gauge transformations. $Q_{d+1}$ acts on d+1 dimensional fields
which are in one to one correspondence with those of the d-dimensional
theory.  $Q_{d+1}$ is identical to
$Q$ when fields are taken at equal values of $x^{d+1}$.  The existence of
the $Q$-exact    action
${\cal I}_{d}$ implies that of an action  ${\cal
I}_{d+1}$ in   $d+1$ dimensions, which is of the
type \cite{nikraba}:
\begin{eqnarray}
\label{nik}
{\cal I}_{d+1}=\int_{M_{d+1}}(\Delta_{d+1}
+\{Q_{d+1},...\}\ )
\end{eqnarray}
  $\Delta_{d+1}$   contains Chern--Simons terms and
  is 
$Q_{d+1}$-invariant but not $Q_{d+1}$-exact. It descends from
the cocycles that one can construct for the cohomology of the $Q$
symmetry  \cite{nikraba}. It  contains terms that are  only
$SO(d)\subset SO(d+1)$ invariant. These general properties have been
analyzed in great details in \cite{nikraba}, and geometrical arguments
have been given for explaining    such a delicate mechanism
that relates theories in d and d+1 dimensions.

  The second term in the r.h.s of Eq.~(\ref{nik}) gives back the  action
${\cal I}_{d}$  we started from,  by standard
dimensional reduction,  up to Chern--Simons terms.
 However, as emphasized in in \cite{nikraba}, one may obtain deformations
of the original  d-dimensional theory.  For instance, 
compactification on a circle provides a  radius $R$ that may be used as  
a parameter for  the deformation,  when one descends from  $d+1$ to $ d $
dimensions.

In our case, we   start from  the   the  8 dimensional
$Q$-exact action that we have introduced in  section 3.
Using   the method of
\cite{nikraba}, we can construct    9-dimensional  fields from the 
 8-dimensional ones  
and 
introduce     the    following
$Q_9$-invariant term in 9 dimensions:
\begin{eqnarray}
\int  \Delta_{9}=\int (
A\wedge dC_3 \wedge dC_3 +B_2^{(-2)} \w dB_2 \wedge dC_3
\nn\\
+dx^9
( \varphi^{(2)}_0   dC_3 \wedge dC_3  + \varphi _1^{(0)}\w dB_2  \w
dC_3 +
\nn\\
\ \ \ \ \ \ \   
\psi_3^{(1)}\w \psi_3^{(1)} \w dA
+dB_2^{(-2)} \w \Psi^{(1)}_2 \wedge \Psi_3^{(1)})  \ )
 \label{cherrr}
\end{eqnarray}
We   can
add to  $ \Delta_{9}$ a $Q_9$-exact action, which is inspired
from the one we have constructed in section 3. Eventually, it is   likely 
that we will  determine  in this way   the 9-dimensional supergravity
action, including Chern-Simons terms.

The first terms in (\ref{cherrr}) are     the dimensional reduction,
modulo d-exact terms,  of the 11-dimensional Chern-Simons term
\begin{eqnarray}
 \Delta_{11}=
C_3\wedge dC_3 \wedge dC_3 
+dx^{11}
( \varphi^{(2)}_2 \w   dC_3 \wedge dC_3  + 
\psi_3^{(1)}\w \psi_3^{(1)} \w dC_3)
\end{eqnarray}
We thus postulate the existence    of a TQFT in 11 dimensions of  a TQFT 
with an action of the following type:
\begin{eqnarray}
\label{mth}
\int (C_3\wedge dC_3 \wedge dC_3  
+dx^{11}
( \varphi^{(2)}_2   dC_3 \wedge dC_3  + 
\psi_3^{(1)}\w \psi_3^{(1)} \w dC_3) +\{ Q_{11},Z_{11} \})
\nn\\
\label{aaa}
\end{eqnarray}
The first terms are  $Q_{11}$-closed but not   $Q_{11}$-exact.  The last term, which is   $Q_{11}$-exact,
determines propagators and regularizes the theory. We   conjecture that
the  existence of the gauge fermion $Z_{11}$ on 11-dimensional manifolds
with
$Spin (7)$ holonomy  follows from that of the TQFT in 8
dimensions. The  action (\ref{aaa})  should be closely  related to that
of 
$N=1,\ D=11$  supergravity on special manifolds. Various types  of
compactifications from 11 to 8 and 7 dimensions may produce interesting
deformations.

Our suggestion  is that the TQFT point of view is likely to  
to  single out the   Chern--Simon part of the     supergravity action in
11 dimensions as a  Q-invariant but not Q-exact terms, while the rest of
the action can be cast in a Q-exact form. The latter part, which includes
the Einstein--Hilbert and Rarita--Schwinger actions, is important to
regularize the theory and
 the path integral;  however, in the topological phase, only the the
Chern--Simons part is truly relevant. 

\section {Conclusion and discussion}

This paper shows that, in eight  dimensions, the TQFT for a metric, two
1-forms, a 2-form and a 3-form reproduces in a twisted form (at least in
the quadratic approximation)  the  $N=2 $  $D=8$   supergravity in a
$Spin(7)$-invariant gravitational background, which is the dimensional
reduction of
 $N=1 $  $D=11$ supergravity. We found the relevant self-duality
equations for defining the TQFT.

 This result  is quite interesting.    In particular, it 
gives an alternative definition for local supersymmetry. It  
  circumvates the
  challenge    of closing the algebra of infinitesimal
transformations of supergravities by mean of  auxiliary fields  and   of   getting  a
group structure   for the   transformations of  supergravity. We now understand   the
fact that the latter    ``close modulo equations of motions" as   a     consequence
of  the   determination of  a TQFT    by enforcing self--duality equations in a
BRST invariant way.

Many dimensional reductions of the TQFT that we have exhibited can be  
thought of.
 We  foresee theories in seven and six  dimensions, where    $ G_2 
\subset Spin(7)  \subset SO(8)$ plays an important role \cite{Bilal}, as
well as  potentially interesting models 
  in 3, 2, 1 and 0 dimensions. In the latter case, there could be 
    a generalization of the DVV  matrix model \cite{dvv} by gravitational
terms,   stemming  from the topological 8-dimensional action. They might
be relevant to the recent works in \cite{vafa}.

 The 8-dimensional gravitational theory  can be also coupled to a non
Abelian Yang--Mills theory, which allows us to consider other types of
models by dimensional reduction, and perhaps, to find   a description of
the octonionic superstring
\cite{stro}.

 A more speculative idea   is that, since   $N=1$ $D=11$ supergravity
seems of a topological origin, it might     encode by dimensional
reduction the  supersymmetric expression of  stochastically  quantized
gauge theories,   which  have  also   the structure of a TQFT,
and      efficiently  regularize many of the problems that one
encounters in the quantization of 4-dimensional Yang--Mills
theories~\cite{lbz}.

We added   the following  remarks. 

The
construction of a TQFT for a 3-form in  eight dimensions can be done
independently of the context of topological gravity.  It still  needs $Spin(7)$
holonomy manifolds, but  the  metric is no longer an independent field.   One   gets 
models of the topological type that  depend  on a gravitino and seem  to  contain  
topological observables.     Such   models are  related to
the``Chern--Simons action" $\int  C_3\w dC_3 $  in seven dimensions.

 We also
suggested  that, on special manifolds, the supergravity action in  eleven
dimensions can be separated into a Q-closed  but  not Q-exact term,  which is the
Chern--Simons part, and a Q-exact  term, which defines the propagation of the fields.

 {\bf Acknowledgments}: I thank all members of RUNHETC, where most of this work has been done. I
am grateful to    M. Douglas, G. Moore, E. Rabinovici, P. Ramond and IM Singer for
interesting discussions. I am indebted to E. Witten for an enlighting  discussion,
which has lead me to add section 4 to this paper.

\end{document}